\documentstyle[epsfig]{aipproc}

\newcommand{\bec}[1] {\begin{equation}\label{#1} }
\newcommand{\eec} {\end{equation} }
\begin{document}

\title{QCD Loop Corrections to Top Production and Decay 
}

\begin{flushright}
UR-1624 \\
ER/40685/961 \\
hep-ph/0012173\\
December 2000 \\
\end{flushright}

\author{ Cosmin Macesanu}
\address{Department of Physics and Astronomy,
University of Rochester,Rochester, New York 14627}

\maketitle

\vspace{-0.5cm}
\begin{abstract}
 The subject of this talk is the computation of virtual QCD
corrections to top production and decay at $e^+ e^-$ colliders. 
We examine the resonant behavior of the amplitudes that dominate
the cross section and discuss the double pole approximation (DPA).  
The theoretical framework is similar to that which has been successfully
applied for QED corrections to $W$ pair production at LEP II.
\end{abstract}

\section*{Introduction}

 An $e^+ e^-$ collider with center-of-mass energy at and above the top 
threshold promises to provide a clean environment in which to perform
precision studies of the top quark.
It is conceivable that at such a machine the study of top can be performed
with a precision similar to that 
achieved in the study of the $W$ boson at LEP II.
This means order \%  (and better) measurements of 
differential cross sections for processes involving the top quark. 
Such a precision in measuring experimental quantities implies the need
for a like precision in our theoretical understanding of these processes.
This in turn requires the inclusion of radiative corrections in our 
predictions.

 Our aim is the computation of QCD corrections to the full production
and decay process (that is, we do not treat the top quarks as on-shell
particles). Therefore, besides corrections to particular subprocesses
(production or decay), we also take into account interference 
(or nonfactorizable) corrections. Similar computations have been performed
for the $W$ pair production process \cite{ddrw}. We discuss
similarities and differences between the two cases.
\section*{Framework}

\begin{figure}
 \begin{minipage}[t]{0.47\linewidth}
   \centering
   \includegraphics[width = 2.5in]{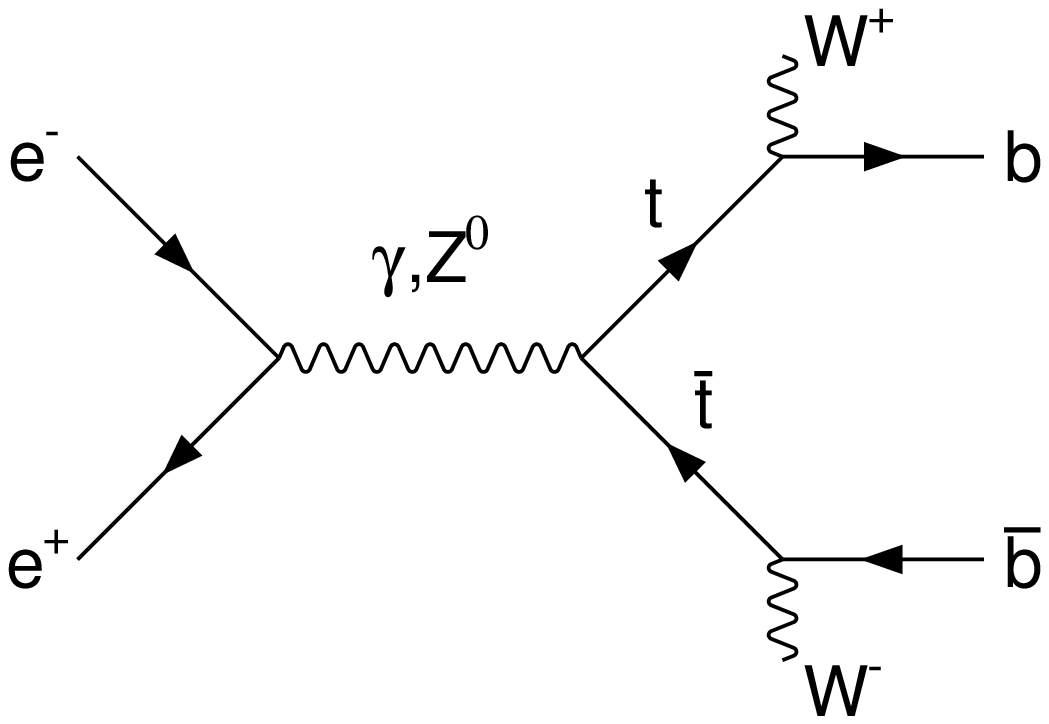}
   \caption{  Tree level diagram  for process (\ref{proc2})}
   \label{fig2}
 \end{minipage}%
 \hspace{0.06\linewidth}%
 \begin{minipage}[t]{0.47\linewidth}
   \centering
   \includegraphics[width = 2.5in]{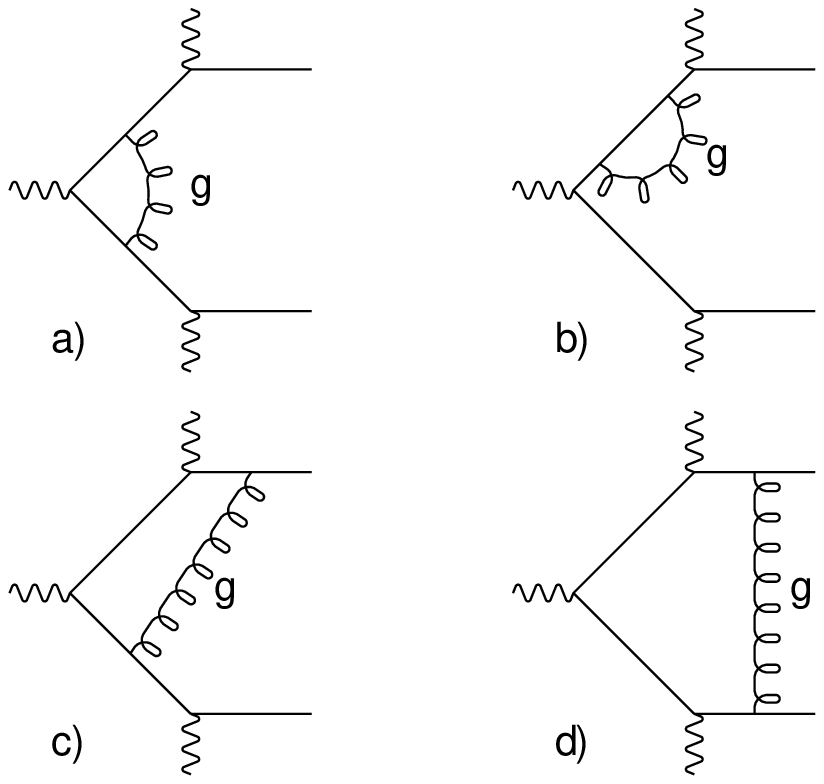}
   \caption{  Virtual corrections to (\ref{proc2}); a) and b) :
corrections to vertex and fermion self energy; c) and d) :
interference-type corrections}
   \label{fig1}
 \end{minipage}
\end{figure}

The process of interest is, at tree level,
\nopagebreak[1]
\bec{proc2}
 e^+ e^- \rightarrow\ t\ \bar{t}\ \rightarrow b\ W^+\ \bar{b}\ W^-
\eec
whose Feynman diagram is represented in Fig. \ref{fig2}. Note that
there are many diagrams contributing to the final state $b\ W^+\ \bar{b}\ W^-$;
however, in the region of the phase space where 
$p_t^2,\ p_{\bar{t}}^2  \approx m_t^2$, (\ref{proc2}) dominates the amplitude
(due to the two resonant top propagators).

Virtual corrections to (\ref{proc2}) can be roughly divided into two classes : 
corections to particular subprocesses -- like vertex and fermion
self-energy diagrams (Figs. 2a and 2b respectively)
 --  and interference-type corrections (as in Figs. 2c, 2d).
For computational purposes, we consider six separate contributions :
\bec{m1amp}
 M_1 = M_{t\bar{t}} + M_{t b} + M_{\bar{t}\bar{b}} + 
M_{b\bar{t}} + M_{t\bar{b}} + M_{b\bar{b}} 
\eec
where the  first three terms correspond to the three vertex corrections 
(in which we have included in a suitable way the fermion self energy
 corrections), and the last
three terms come from the interference diagrams. The exact evaluation
of these partial amplitudes is a difficult task,
especially for the interference part.
 However, we don't need the exact results; 
these amplitudes are also dominated by terms 
which are enhanced by two resonant top propagators. Hence, we will restrict
ourselves to the computation of these doubly resonant terms (double pole
approximation). 

 As an example of how this approximation works, let's look at the 
production-top decay interference amplitude
\footnote{ here, $ \bar{m}_t^2 =  m_t^2 - i m_t \Gamma_t$} :
$$ M_{b\bar{t}} =
 \bar{u}(b) \left[ (-i g_s^2)
 \int \frac{d^4 k}{2\pi^4}\ \frac{1}{k^2}\
\gamma^\mu \ \frac{\hat{p}_b-\hat{k} + m_b}{(p_b-k)^2 - m_b^2}\
\hat{\epsilon}_{W^+} \ \frac{\hat{p}_t-\hat{k} + m_t}{(p_t-k)^2 - \bar{m}_t^2}\
\hat{a}\ \right.
$$
\bec{mbbart}
\left.
\frac{-\hat{p}_{\bar{t}} - \hat{k} + m_t}{(p_{\bar{t}}+k)^2 - \bar{m}_t^2}\
\gamma_\mu \right] \ 
\frac{-\hat{p}_{\bar{t}} + m_t}{p_{\bar{t}}^2 - \bar{m}_t^2}\
\hat{\epsilon}_{W^-} \ v(\bar{b})
\eec
The doubly resonant terms can be extracted with the help of a simple
observation: if the virtual gluon in the loop is hard, then 
the quantity in  brackets will be well behaved, and the overall
resonant behavior for this amplitude is $1/(p_{\bar{t}}^2 - \bar{m}_t^2)$.
This means that the doubly resonant terms are entirely due to soft 
virtual gluons. Therefore, we can use the extended soft gluon
approximation (ESGA), which means neglecting the $\hat{k}$ terms in 
the the numerator of (\ref{mbbart}) (see also \cite{ddr},\cite{lynne}).
Then, with the help of some standard
Dirac algebra manipulations, we obtain :
\bec{mbbtr}
 M_{b\bar{t}}(DPA+ESGA) = 
\frac{\alpha_s}{4\pi} M_0 \ * \ (-4 p_b p_{\bar{t}})( p_t^2-\bar{m}_t^2 ) \ 
D^0_{b\bar{t}}
\eec
where $M_0$ is the tree level matrix element, and $D^0_{b\bar{t}}$ is the
scalar 4-point function corresponding to the production-top decay
interference diagram
\footnote{
$$
D^0_{b\bar{t}} = \int \frac{d^4 k}{i\pi^2}\ \frac{1}{k^2 + i \epsilon}\ 
\frac{1}{k^2-2kp_b}\
\frac{1}{(p_t-k)^2 - \bar{m}_t^2}\ \frac{1}{(p_{\bar{t}}+k)^2 - \bar{m}_t^2}\ 
$$
}.

Let's take a look at the resonant behavior of $M_{b\bar{t}}$.
The $( p_t^2-\bar{m}_t^2 )$
factor in the above equation cancels a pole in $M_0$, and the $D_0$ 
function has a logarithmic singularity when the $\bar{t}$ goes on-shell. 
Therefore, the resonant behavior of $M_{b\bar{t}}$ is of type 
$pole \times log$ rather than double pole. Similar results and
behavior are obtained for the other interference terms.

The results for the purely non-factorizable corrections (another name for the
interference diagrams 2c, 2d)
for the $t\ \bar{t}$ production and decay 
process are completely
analogous to the results obtained for the same type of diagrams in 
the W pair production case \cite{ddr}.
However, this similarity does not hold
for the vertex corrections. Consider, for example, the vectorial part of 
the $t\ \bar{t}$ production vertex :
$$ \delta \Gamma^{\mu}_V = \frac{\alpha_s}{4 \pi} \
\int \frac{d^4 k}{i \pi^2}\ \frac{1}{k^2}\ \gamma^{\nu}\
\frac{\hat{p}_t - \hat{k} + m_t}{(p_t - k)^2 - \bar{m}_t^2}\
\gamma^{\mu}C_V\
\frac{-\hat{p}_{\bar{t}} - \hat{k} + m_t}{(p_{\bar{t}} + k)^2 - \bar{m}_t^2}\
\gamma_{\nu}
$$
 Upon integration, we can express the result in terms of 8 form factors :
$$
\delta \Gamma^{\mu}_V = \frac{\alpha_s}{4 \pi} C_V\
\left[ {\gamma^{\mu}} {F_2} + { (\hat{p}_t - m_t)\gamma^{\mu}} {F_4}
 + {\gamma^{\mu}(\hat{p}_{\bar{t}} + m_t)} {F_6} +
 {(\hat{p}_t - m_t)\gamma^{\mu}(\hat{p}_{\bar{t}} + m_t)} {F_8}
\right.
$$
\bec{mttb1}
 \left.  + 
{(p_t - p_{\bar{t}})^{\mu}} {F_1} +  
(\hat{p}_t - m_t) (p_t - p_{\bar{t}})^{\mu} F_3   +\ldots \right]
\eec
In the on-shell case, only the $F_2$ (electric dipole) and $F_1$ (magnetic dipole momentum) form factors contribute. We might expect that in the 
double pole approximation we can drop the other terms, too, since
the $ (\hat{p}_t - m_t) $ terms will cancel a pole (or both) in the amplitude.
However, the form factors themselves have a logarithmic resonant behavior 
when either one particle (for the decay vertices) or both (for the production
vertex) go on-shell.
It follows that in expression 
(\ref{mttb1})
we have to keep the terms which contain $F_1$ to $F_6$  (we can drop the 
 $F_7$ and $F_8$ terms), and we do not have factorization 
(virtual corrections being proportional to the tree level matrix element)
anymore. This is 
different from what happens in the $W$ pair production process,
where in DPA factorization holds even for the vertex corrections. This
difference is due to the fact that in our process the intermediate particles
are fermions, and not bosons.

An issue which needs some attention is the gauge invariance of our result.
A complete computation, taking into account all the diagrams, including
the non-resonant ones, would give a gauge invariant answer; however, 
by restricting ourselves to a subset of diagrams, we may lose this property.
In the W pair production case, gauge invariance in DPA is guaranteed by 
the fact that the off-shell corrections are proportional to the on-shell
result. This is no longer true here; however, we have checked that our 
results are gauge invariant in the approximation used (that is, up to
singly resonant terms). 

As the last topic, we will mention only briefly
the treatment of infrared singularities. In 
the computation of the virtual corrections, the only place they appear is
in the self-energies of the $b$ and $\bar{b}$ quarks and in the decay-decay
interference diagram 2d. Once we take into account the soft gluon radiation,
these virtual singularities will cancel against the ones coming from 
real gluon radiation from the $b$ and $\bar{b}$ quarks.

\section*{Conclusions}

The subject of this paper concerns QCD corrections to top production 
and decay at a linear collider. We have discussed in some detail 
the implementation of the double pole approximation in the evaluation
of the dominant diagrams which contribute to this process.
Similarities and differences to the computation 
of QED radiative corrections to $W$ pair production at LEP have been
mentioned.

This is work in progress, and we expect to have numerical results soon. 
The final aim of our work is a complete next to leading order QCD 
computation of top production and decay at an $e^+ e^-$ collider. 

I wish to thank my adviser Prof. Lynne Orr and Dr. Doreen Wackeroth
for support offered and many helpful discussions on the 
subject of virtual corrections.

\end{document}